\documentclass[11pt]{article}

\nonstopmode
\usepackage{natbib,hypernat}
\usepackage{graphicx,wrapfig,graphics} 
\usepackage[nodayofweek]{datetime}
\usepackage[footnotesize]{caption}
\usepackage{array}
\usepackage{color}
\usepackage[utf8]{inputenc}
\usepackage{algorithmic}
\usepackage[ruled]{algorithm}
\usepackage{parskip}
\usepackage{listings}

\usepackage{amsmath}
\usepackage{amsfonts}
\usepackage{amssymb}
\usepackage{stmaryrd}

\usepackage{float}
\floatstyle{plain}
\newfloat{proglisting}{tpb}{listings}
\floatname{proglisting}{Listing}

\usepackage{tikz}
\usetikzlibrary{positioning}
\usetikzlibrary{calc}
\usetikzlibrary{chains}

\pgfrealjobname{main}

\newcommand{\doi}[1]{\href{http://dx.doi.org/#1}{doi: #1}}

\usepackage{hyperref}
\usepackage{subfigure}

\title{Solving Wave~Equations on Unstructured~Geometries}

\author{
Andreas Klöckner\footnote{%
  Courant Institute of Mathematical Sciences,
  New York University,
  New York, NY 10012}\,\
\and
Timothy Warburton\footnote{%
  Department of Computational and Applied Mathematics,
  Rice University,
  Houston, TX 77005}\,\
\and
Jan S. Hesthaven\footnote{%
  Division of Applied Mathematics,
  Brown University,
  Providence, RI 02912}
}

\definecolor{darkgreen}{rgb}{0,0.4,0}

\newcommand{\ednote}[1]{}
\newcommand{\authnote}[1]{}

\newcommand{\assign}{:=}

\newcommand{\I}{\mathsf{I}}
\newcommand{\D}{\mathsf{D}}

\newcommand{\mathd}{\,\mathrm{d}}
\newenvironment{myfigure}{\begin{figure}\centering}{\end{figure}}

\colorlet{codeback}{gray!20}

\begin{document}
\maketitle

%
%
%
%



\ednote{A paragraph with no section heading to introduce the chapter.}

Waves are all around us--be it in the form of sound, electromagnetic
radiation, water waves, or earthquakes. Their study is an important
basic tool across engineering and science disciplines. Every wave
solver serving the computational study of waves meets a trade-off of
two figures of merit--its computational speed and its accuracy.
Discontinuous Galerkin (DG) methods fall on the high-accuracy end of
this spectrum. Fortuitously, their computational structure is so
ideally suited to GPUs that they also achieve very high computational
speeds. In other words, the use of DG methods on GPUs significantly
lowers the cost of obtaining accurate solutions. This article aims to
give the reader an easy on-ramp to the use of this technology, based
on a sample implementation which demonstrates a highly accurate,
GPU-capable, real-time visualizing finite element solver in about 1500
lines of code.

\section{Introduction, Problem Statement, and Context}
\ednote{This section should include a description of the general problem 
being solved and how this technology helps to improve the state of the art,
but do so at two levels, one targeting non-domain-experts, and others for 
readers in this field.}

At the beginning of our journey into high-performance, highly accurate
time-domain wave solvers, let us briefly illustrate by a few examples
how common the task of simulating wave phenomena is across many
disciplines of science and engineering, and how accuracy figures into
each of these application areas. Consider the following examples:

\begin{itemize}

  \item An engineer needs to understand the time-domain response of an
  oscillating structure such as an accelerator cavity.  Real-life
  measurement of the desired properties is extremely costly, if it is
  possible at all. Accuracy is important because wrong results may
  lead to wrong conclusions.

  \item A seismic engineer has time-domain data from a sounding using
  geophones and needs to model an underground structure, characterized
  by different wave propagation speeds. Doing so requires a solver for
  the `forward problem', i.e. a code that, given data about the
  location of the sources and wave propagation speeds throughout the
  underground domain, can model the propagation of the waves.
  Accuracy is important because these simulations often inform
  potentially very costly enterprises such as drilling or mining.

  \item An electrical engineer wants to model the stealth properties
  of a new airplane involving complicated nonlinear materials.
  Physical prototyping is expensive, and accurate predictions of
  scattering properties help minimize its necessity.

\end{itemize}

In the field of time-domain wave simulation, the main competitors of
the discontinuous Galerkin method include finite-difference,
finite-volume and continuous finite-element methods.  In a nutshell,
finite-difference solvers have trouble representing complicated
geometric boundaries, finite-volume methods become very difficult (and
very expensive) to implement at a high order of accuracy\footnote{The 
order of accuracy refers to the power with which the error decreases
as the discretization is refined--for example, if the distance between
neighboring mesh points is halved, a fourth-order scheme would
decrease the error by a factor of sixteen.}, and
continuous finite-element methods typically assemble large, sparse
matrices, whose application to a vector is necessarily memory-bound
and thus unable to make use of the massive compute bandwidth available
on a GPU.

In addition, while finite-difference methods have relatively benign
implementation properties on GPUs
\cite{micikevicius_3d_2009,cohen_opencurrent_2009}, we will see that
the computational structure of DG methods is even better suited to GPU
implementation at high accuracy because they largely avoid the wide
``halo'' of outside values that must be fetched in order to apply a
large (high-order) stencil to three-dimensional volume data.

This chapter complements an article \citep{kloeckner_nodal_2009} which
we have recently published that, in its spirit, is probably more like
the other chapters in this volume in that it exposes all the
technicalities and tricks that have enabled us to demonstrate
high-speed DG on the GPU. To avoid redundancy between
\citep{kloeckner_nodal_2009} and this chapter, we have instead chosen
to focus our treatment here on easing a prospective user's entry into
using our technology. While \citep{kloeckner_nodal_2009} is very
technical and not entirely suited as an introduction to the subject,
in this chapter we will be applying a number of simplifications 
to facilitate understanding and promote ease-of-use.

\section{Core Method}
\ednote{Concise overview of underlying methods used in this solution.
One of the key objectives of this section should be to provide enough 
information to nondomain experts to allow them to make a determination 
whether they should read the rest of the chapter. 
In particular, readers not from your field will try to see if the 
techniques in your chapter will be applicable to their problems. 
It is therefore important to have a good illustration that allows 
readers from other fields to quickly determine if the problem being
solved is similar to the ones they are trying to solve in their own fields.}

Discontinuous Galerkin (DG) methods for the numerical solution of
partial differential equations have enjoyed considerable success
because they are both flexible and robust: They allow arbitrary
unstructured geometries and easy control of accuracy without
compromising simulation stability.  Lately, another property of DG has
been growing in importance: The majority of a DG operator is applied
in an element-local way, with weak penalty-based element-to-element
coupling.

The resulting locality in memory access is one of the factors that
enables DG to run on off-the-shelf, massively parallel graphics
processors (GPUs).  In addition, DG's high-order nature lets it
require fewer data points per represented wavelength and hence fewer
memory accesses, in exchange for higher arithmetic intensity.  Both of
these factors work significantly in favor of a GPU implementation of
DG.

Readers wishing a deeper introduction to the numerical method are
referred to the introductory textbook \cite{hesthaven_nodal_2007}.

\section{Algorithms, Implementations, and Evaluations}
\ednote{This is the core of the chapter. This can be a multi-subsection section
It presents key insights from algorithm implementations and
evaluations along the way. Each subsection can give the algorithm, 
implementation, and measured benefit of a version of your solution. 
The section should be written to maximize the likelihood that someone 
can reproduce your success.}

\subsection{Background Material }
\subsubsection{A Precise Mathematical Problem Statement}

Discontinuous Galerkin methods are most often used to solve
hyperbolic systems of conservation laws in the time domain. This
rather general class of partial differential equation (PDE) can be
written in the form
\begin{equation}
\label{eq:claw}
\frac{\partial q}{\partial t}+\nabla_x \cdot F(q) = f.
\end{equation}
DG methods generally solve the \emph{initial boundary value problems}
(IBVPs) of these equations on a bounded domain $\Omega$. This means
that in addition to the PDE \eqref{eq:claw}, one needs to specify the
finite geometry of interest, an initial value of the solution $q$ at
an initial time $T_0$ (which we will assume to be zero) as well
as which (potentially time-dependent) conditions prevail at the
boundary $\partial \Omega$ of the domain. In addition, source terms
may be present. These are represented in \eqref{eq:claw} by $f$.

Classes of partial differential equations more general than
\eqref{eq:claw}, such as parabolic and elliptic equations, can be
solved using DG methods. In this chapter, we will focus on
hyperbolic equations, and for the sake of exposition, on one
particularly important example of these equations, the second-order
wave equation in two dimensions.  To emphasize the equation's
grounding in reality, we will cast this equation as (the
transverse-magnetic version of) the linear, isotropic,
constant-coefficient Maxwell's equations in two dimensions and show
the method's development by its example. The equation itself is given
by
\begin{subequations}
\label{eq:tm-maxwells}
\begin{align}
  \label{eq:tm-maxwells-hx}
  0&=\mu \frac{\partial H_x}{\partial t} 
  + \frac{\partial E_z}{\partial y},\\
  \label{eq:tm-maxwells-hy}
  0&=\mu \frac{\partial H_y}{\partial t} 
  - \frac{\partial E_z}{\partial y},\\
  \label{eq:tm-maxwells-ez}
  0&=\epsilon \frac{\partial E_z}{\partial t} 
  - \frac{\partial H_y}{\partial x}
  + \frac{\partial H_x}{\partial y}.
\end{align}
\end{subequations}
One easily verifies that this equation can be rewritten into the more 
well-known second order form of the wave equation,
\[ 
  \frac{\partial^2 E_z}{\partial t^2}
  =c^2 \triangle E_z
\]
with $c^{-2}=\epsilon \mu$.  For simplicity, we may assume 
$c=\epsilon=\mu=1$. Together with an initial condition
as well as perfectly electrically conducting (PEC) boundary
condition
\[
  E_z(x,t)=0\quad\text{on $\partial \Omega$}.
\]
Observe that no value is prescribed for the magnetic fields $H_x$,
$H_y$, which we leave to obey \emph{natural boundary conditions}.
In terms of the second-order wave equation, PEC corresponds to a
Dirichlet boundary.

\subsubsection{Construction of the Method}
\label{sec:dg-method-construction}

To begin the discretization of \eqref{eq:tm-maxwells}, we assume
that the domain $\Omega$ is polyhedral, so that it may be represented
as a union $\Omega = \biguplus_{k=1}^K \D_k \subset \mathbb R^2$ consisting of
disjoint, straight-sided, face-conforming triangles $\D_k$.

We demonstrate the construction of the method by the example
of equation \eqref{eq:tm-maxwells-ez}. We begin by multiplying
\eqref{eq:tm-maxwells-ez} with a test function $\phi$ and integrating
over the element $\D_k$:
\begin{align*}
  0 &= \int_{\D_k} \frac{\partial E_z}{\partial t} \phi \mathd V
  - \int_{\D_k} \frac{\partial H_y}{\partial x} \phi \mathd V
  + \int_{\D_k} \frac{\partial H_x}{\partial y} \phi \mathd V\\
  &= \int_{\D_k} \frac{\partial E_z}{\partial t} \phi \mathd V
  + \int_{\D_k} \nabla_{(x,y)} \cdot 
  \underbrace{(-H_y, H_x)^T}_{F=} \phi \mathd V.
\end{align*}
Observe that the vector-valued $F$ indicated here assumes the role of
the flux $F$ in \eqref{eq:claw}.  Integration by parts yields
\begin{equation}
  \label{eq:dg-weak}
  0=
  \int_{\D_k} \frac{\partial E_z}{\partial t} \phi \mathd V
  - \int_{\D_k}  (-H_y, H_x)^T \cdot \nabla_{(x,y)} \phi \mathd V
  + \int_{\partial \D_k}  \hat n \cdot  (-H_y, H_x)^T \phi \mathd S,
\end{equation}
where $\hat n$ is the unit normal to $\partial \Omega$. Now a key
feature of the method enters. Because no continuity is enforced on
$H_x$ and H$_y$ between $\D_k$ and its neighbors, the value of $H_x$
and $H_y$ on the boundary is not uniquely determined. For now, we will
record this fact by a superscript asterisk, denote these chosen values
the \emph{numerical flux}, and leave a determination of what value
should be used for later.

To revert the so-called \emph{weak form} \eqref{eq:dg-weak} to a shape
more closely resembling the original equation
\eqref{eq:tm-maxwells-ez}, we integrate by parts again, obtaining the
so-called \emph{strong form}
\begin{multline}
  \label{eq:dg-strong}
  0=\int_{\D_k} \frac{\partial E_z}{\partial t} \phi \mathd V
  + \int_{\D_k} \nabla_{(x,y)} \cdot (-H_y, H_x)^T \phi \mathd V
  \\
  - \int_{\partial \D_k}  \hat n \cdot  
  (-(H_y-H_y^*), H_x-H_x^*)^T \phi \mathd S
\end{multline}
where we carefully observe that the boundary term obtained 
in the last step has stayed in place.

To determine the values of $H^*$, we note that in many cases a simple
average across neighboring faces, i.e.  $H^*:=(H^+ + H^-)/2$ leads to a
stable and accurate numerical method, where $H^-$ denotes the values
on the local face.  This is termed a \emph{central flux}. We choose
a more dissipative (but less noisy) \emph{upwind flux}
\cite{mohammadian_computation_1991}, given by
\begin{equation}
\label{eq:max2d-numflux}
\hat n \cdot (F-F^*)
=\begin{bmatrix}
  \hat n_y [E_z] + \alpha \hat n_x(\hat n_x [H_x] + \hat n_y [H_y] - [H_x])\\
  -\hat n_x [E_z] + \alpha \hat n_y(\hat n_x [H_x] + \hat n_y [H_y] - [H_y])\\
  \hat n_y [H_x] -\hat n_x [H_y] -\alpha [E_z]
\end{bmatrix}.
\end{equation}
The value to be used for $\hat n \cdot (-(H_y-H_y^*), H_x-H_x^*)^T$ in
\eqref{eq:dg-strong} can be read from the third entry of the right
hand side of \eqref{eq:max2d-numflux}, and the first two entries apply
to equations \eqref{eq:tm-maxwells-hx} and \eqref{eq:tm-maxwells-hy}.
We have used the common notation $[q]=q^--q^+$ for the inter-element
jumps. $\alpha$ is a parameter, commonly chosen as $1$.  Obviously,
$\alpha=0$ recovers a central flux.

We expand $E$, $H$, and $\phi$ into a basis of $N_p$ Lagrange
interpolation polynomials $l_i$ spanning the space $P^N$ of
polynomials of total degree $N$, where the Lagrange interpolation
points are purposefully chosen for numerical stability
\cite{warburton_explicit_2006}. Substituting the expansions
into \eqref{eq:dg-strong} combined with \eqref{eq:max2d-numflux}
yields a numerical scheme that is discrete in space, but not
yet in time.

\subsubsection{Implementation Aspects}

To actually implement this scheme, we express
\eqref{eq:dg-strong} in matrix form. To do so, first 
note that in our setting, each element $\D_k\subset\Omega$ can
be obtained by an affine map $\Psi(r,s)=A_k(r,s)^T+b_k$
from a reference element $\I$.
Now define the mass matrix
\[
\mathcal M_{i j}^k \assign \int_{\D_k} l_i l_j \mathd V
=|A_k|\mathcal M\assign |A_k| \int_{\I} l_i l_j \mathd V.
\]
$|A_k|$ is the determinant of the matrix $A_k$.  Also let $\mathcal
D^{\partial \nu}$ be the matrix that realizes polynomial
differentiation along the reference element's $\nu$th axis in Lagrange
coefficients. Polynomial differentiation along \emph{global}
coordinates is realized as a linear combination of these local
differentiation matrices, according to, e.g.,
\begin{equation}
  \label{eq:dg-diff-local-to-global}
  \mathcal D^{k,\partial x} 
  = (A_k^{-1})_{11} \mathcal D^{\partial 1}
  +(A_k^{-1})_{12} \mathcal D^{\partial 2}.
\end{equation}
This allows us to express an implementation of
the volume part of \eqref{eq:dg-strong}:
\begin{multline}
  \label{eq:dg-strong-matrix}
  0=|A_k| \mathcal M  \frac{\partial (E_z)_N}{\partial t}
  + |A_k| \mathcal M (
  \mathcal D^{k,\partial x} (-H_y)_N+ 
  \mathcal D^{k,\partial y} (H_x)_N)
  \\
  - \int_{\partial \D_k}  \hat n \cdot  
  (-(H_y-H_y^*), H_x-H_x^*)^T \phi \mathd S.
\end{multline}
 For numerical
stability at increasing $N$, the matrices $\mathcal M^k$ and $\mathcal
D^{\partial \nu}$ are computed by ways of orthogonal polynomials on
the triangle
\citep{koornwinder_two-variable_1975,dubiner_spectral_1991}.

To implement the surface terms, define the surface mass matrix
for a single face $\Gamma$ of the reference triangle $\I$:
\[
M_{i j}^{\Gamma} \assign \int_{\Gamma \subset \partial \I} 
l_i l_j \mathd S.
\]
Suppose we compute values of 
$\hat n \cdot (F-F^*)=\hat n \cdot  (-(H_y-H_y^*), H_x-H_x^*)^T$ 
along all faces and concatenate these
into one vector. Then the sum over all facial integrals may be
computed through a carefully assembled matrix:
\begin{multline}
\label{eq:dg-lift}
\int_{\partial \D_k}  \hat n \cdot  
  (F-F^*) \phi \mathd S
=\\
\text{
\begin{minipage}{3.4cm}
\beginpgfgraphicnamed{multi-face-mass-matrix}
\begin{tikzpicture}[
  densemat/.style={fill=gray!30},
  ]
  \draw (2,-1) rectangle +(3,2) ;
  %
  \draw [densemat] (2,1) rectangle +(1,-1) ;
  \foreach \ys/\ye in {0.1/0.3, -0.25/-0.7, 0.6/0.8}
  { \draw [densemat] (3,\ys) rectangle (4,\ye) ; }
  \foreach \ys/\ye in {0.4/0.1, -0.1/-0.3, -0.9/-1, 1/0.9}
  { \draw [densemat] (4,\ys) rectangle (5,\ye) ; }
  \node at (2.5,0.5) {$M^{\Gamma_1}$} ;
  \node at (3.5,-0.475) {$M^{\Gamma_2}$} ;
  \node at (4.5,0.25) {$M^{\Gamma_3}$} ;
\end{tikzpicture}
\endpgfgraphicnamed
\end{minipage}
}
\bigg(
J_1 \hat n \cdot  (F-F^*) | _{\Gamma_1}
\bigg|
\cdots
\bigg|
J_3 \hat n \cdot  (F-F^*) | _{\Gamma_3}
\bigg).
\end{multline}
We denote this matrix $\mathcal M^{\partial \I}$ and the vector to which we
are applying it $\mathbf{f}^k$.  The factors $J_n$ are the
determinants of the affine maps parametrizing the faces of $\D_k$
with respect to the faces of $\I$.

Returning to \eqref{eq:dg-strong-matrix}, we left-multiply by
$|A_k|^{-1} \mathcal M^{-1}$ to obtain
\begin{equation}
  \label{eq:dg-strong-explicit}
  0=\frac{\partial (E_z)_N}{\partial t}
  + (
  \mathcal D^{k,\partial x} (-H_y)_N+ 
  \mathcal D^{k,\partial x} (H_x)_N)
  \\
  - |A_k|^{-1} \mathcal M^{-1} \mathcal M^{\partial \I} \mathbf{f}^k
\end{equation}
Despite all the machinery involved, \eqref{eq:dg-strong-explicit} is
strikingly simple, consisting of three data-local element-wise
matrix-vector multiplications (two differentiations, one combined face
mass matrix) and a surface flux exchange term. A view of the flow of
data is provided by Figure \ref{fig:dg-subtasks}.

\begin{myfigure}
  \centering
  \beginpgfgraphicnamed{dg-operator-decomposition}
  \begin{tikzpicture}[
    txt/.style={text height=1.5ex, text depth=0.25ex},
    operation/.style={rectangle,draw,minimum height=5ex,txt},
    localop/.style={operation,line width=2pt,txt},
    data/.style={circle,draw,minimum size=7ex,txt},
    ]

    \node [data] (state) { $u^k$ } ;
    \node [operation] (gather) [above right=0.1cm and 0.5cm of state] { Flux Gather } ;
    \node [localop] (lift) [right=1cm of gather] { Surface Integration  } ;
    \node [localop] (fu) [below right=0.1cm and 0.5cm of state] { $F(u^k)$ } ;
    \node [localop] (diff) [right=1cm of fu] { Local Differentiation} ;
    \node [data] (rhs) [below right=0.1cm and 0.5cm of lift] { $\partial_t u^k$ } ;
    \draw [->] (state) |- (gather) ;
    \draw [->] (gather) |- (lift) ;
    \draw [->] (lift) -| (rhs) ;

    \draw [->] (state) |- (fu) ;
    \draw [->] (fu) |- (diff) ;
    \draw [->] (diff) -| (rhs) ;
  \end{tikzpicture}
  \endpgfgraphicnamed
  \caption{Decomposition of a DG operator into
  subtasks. Element-local operations are highlighted with a bold outline.}
  \label{fig:dg-subtasks}
\end{myfigure}

Even better, the time derivative
$\frac{\partial (E_z)_N}{\partial t}$ occurs on its own, making it
possible to use simple, explicit Runge-Kutta methods for integration
in time.

\subsection{A Minimal Implementation}
\label{sec:dg-minimal-implementation}

After this very quick (but mostly self-contained) introduction to
discontinuous Galerkin methods, we will now discuss how
\eqref{eq:dg-strong-explicit} and its analogous extension to
\eqref{eq:tm-maxwells-hx} and \eqref{eq:tm-maxwells-hy} may be brought
onto the GPU to form a solver for the 2D TM variant of Maxwell's
equations.

\subsubsection{Introduction}
To make the discussion both more tangible and easier to follow, we
have created a simple implementation of the ideas presented here.
This implementation may be downloaded from the URL
\url{http://tiker.net/gcg-dg-code-download}.
As improvements are made, the code at this address may change from
time to time.  The source code may also be browsed on-line at
\url{http://tiker.net/gcg-dg-code-browse}.

We will begin by briefly discussing the construction of this package.
The solver is written in Python. We feel that this allows for clearer
code that, in both notation and structure, closely resembles the
MATLAB codes of \cite{hesthaven_nodal_2007}, and yet allows a simple
and concise GPU implementation to be added using, in this case,
PyOpenCL (PyCUDA, whose use is demonstrated in another chapter of this
volume, would have been another obviously possible implementation
choice). In addition, the solver is designed for clarity, not peak
performance.  What we mean here is that the compute kernels we show
are rather simple and lack a few performance optimizations. The
solver's performance is not related to its implementation language.
High-performance GPU codes can easily be constructed using PyOpenCL
(and PyCUDA), which is demonstrated below and in a number of other
chapters of this volume. Finally, we would like to remark that the
kernels as shown below are optimized for Nvidia GPUs and run well on
chips ranging from the G80 to the GF100.

In discussing the solver, we focus on the performance-relevant kernels
running on the GPU. There are other, significant parts of the solver
that deal with preparation and administrative issues such as mesh
connectivity and polynomial approximation. These parts are obviously
also important to the success of the method, but they are beyond the
scope of this chapter. The interested reader may find them explained
more fully in the introductory book \cite{hesthaven_nodal_2007}.  Once
we have discussed the functioning of the basic solver, we will
describe any additional steps that may be taken to improve
performance. Lastly, we will discuss a set of features that may be
added to this rather bare implementation to make it more useful.
In the next chapter, we close by showing performance numbers first for
this solver, and then for our production solver, which is  a more
complete implementation of the ideas to follow.

\subsubsection{Computing the Volume Contribution}
\label{sec:dg-volume-kernel}

First in our examination of implementation features is what we call
the ``volume kernel'', which achieves element-local differentiation as
described in \eqref{eq:dg-diff-local-to-global} and
\eqref{eq:dg-strong-matrix}. The key parts of the kernel's OpenCL C
source are given in Listing \ref{lst:pydgeon-volume-kernel}.

\begin{proglisting}
\begin{lstlisting}[linerange=start_vol_kernel-end]
# Pydgeon - the Python DG Environment
# (C) 2009, 2010 Tim Warburton, Xueyu Zhu, Andreas Kloeckner
#
# This program is free software: you can redistribute it and/or modify
# it under the terms of the GNU General Public License as published by
# the Free Software Foundation, either version 3 of the License, or
# (at your option) any later version.
#
# This program is distributed in the hope that it will be useful,
# but WITHOUT ANY WARRANTY; without even the implied warranty of
# MERCHANTABILITY or FITNESS FOR A PARTICULAR PURPOSE.  See the
# GNU General Public License for more details.
#
# You should have received a copy of the GNU General Public License
# along with this program.  If not, see <http://www.gnu.org/licenses/>.



from __future__ import division

import numpy as np




# {{{ CPU

def MaxwellRHS2D(discr, Hx, Hy, Ez):
    """Evaluate RHS flux in 2D Maxwell TM form."""

    from pydgeon.tools import eldot

    d = discr
    l = discr.ldis

    # Define field differences at faces
    vmapM = d.vmapM.reshape(d.K, -1)
    vmapP = d.vmapP.reshape(d.K, -1)

    dHx = Hx.flat[vmapM]-Hx.flat[vmapP]
    dHy = Hy.flat[vmapM]-Hy.flat[vmapP]
    dEz = Ez.flat[vmapM]-Ez.flat[vmapP]

    # Impose reflective boundary conditions (Ez+ = -Ez-)
    dHx.flat[d.mapB] = 0
    dHy.flat[d.mapB] = 0
    dEz.flat[d.mapB] = 2*Ez.flat[d.vmapB]

    # evaluate upwind fluxes
    alpha  = 1.0
    ndotdH =  d.nx*dHx + d.ny*dHy
    fluxHx =  d.ny*dEz + alpha*(ndotdH*d.nx-dHx)
    fluxHy = -d.nx*dEz + alpha*(ndotdH*d.ny-dHy)
    fluxEz = -d.nx*dHy + d.ny*dHx - alpha*dEz

    # local derivatives of fields
    Ezx, Ezy = d.grad(Ez)
    CuHx, CuHy, CuHz = d.curl(Hx, Hy,0)

    # compute right hand sides of the PDE's
    rhsHx = -Ezy  + eldot(l.LIFT, (d.Fscale*fluxHx))/2.0
    rhsHy =  Ezx  + eldot(l.LIFT, (d.Fscale*fluxHy))/2.0
    rhsEz =  CuHz + eldot(l.LIFT, (d.Fscale*fluxEz))/2.0
    return rhsHx, rhsHy, rhsEz

# }}}

# {{{ OpenCL

# {{{ kernels

MAXWELL2D_VOLUME_KERNEL = """
#define p_N %(N)d
#define p_Np %(Np)d
#define BSIZE %(BSIZE)d

__kernel void MaxwellsVolume2d(int K,
   __read_only __global float *g_Hx,
   __read_only __global float *g_Hy,
   __read_only __global float *g_Ez,
   __global float *g_rhsHx,
   __global float *g_rhsHy,
   __global float *g_rhsEz,
   __read_only __global image2d_t i_DrDs,
   __read_only __global float *g_vgeo)
{
  const sampler_t samp =
    CLK_NORMALIZED_COORDS_FALSE
    | CLK_ADDRESS_CLAMP
    | CLK_FILTER_NEAREST;

  __local float l_Hx[BSIZE];
  __local float l_Hy[BSIZE];
  __local float l_Ez[BSIZE];

  /* LOCKED IN to using Np work items per group */
// start_vol_kernel
  const int n = get_local_id(0);
  const int k = get_group_id(0);

  int m = n+k*BSIZE;
  int id = n;

  l_Hx[id] = g_Hx[m];
  l_Hy[id] = g_Hy[m];
  l_Ez[id] = g_Ez[m];

  barrier(CLK_LOCAL_MEM_FENCE);

  float dHxdr=0,dHxds=0;
  float dHydr=0,dHyds=0;
  float dEzdr=0,dEzds=0;

  float Q;
  for(m=0; m<p_Np; ++m)
  {
    float4 D = read_imagef(i_DrDs, samp, (int2)(n, m));

    Q = l_Hx[m]; dHxdr += D.x*Q; dHxds += D.y*Q;
    Q = l_Hy[m]; dHydr += D.x*Q; dHyds += D.y*Q;
    Q = l_Ez[m]; dEzdr += D.x*Q; dEzds += D.y*Q;
  }

  const float drdx = g_vgeo[0+4*k];
  const float drdy = g_vgeo[1+4*k];
  const float dsdx = g_vgeo[2+4*k];
  const float dsdy = g_vgeo[3+4*k];

  m = n+BSIZE*k;
  g_rhsHx[m] = -(drdy*dEzdr+dsdy*dEzds);
  g_rhsHy[m] =  (drdx*dEzdr+dsdx*dEzds);
  g_rhsEz[m] =  (drdx*dHydr+dsdx*dHyds 
    - drdy*dHxdr-dsdy*dHxds);
// end
}
"""

MAXWELL2D_SURFACE_KERNEL = """
#define p_N %(N)d
#define p_Np %(Np)d
#define p_Nfp %(Nfp)d
#define p_Nfaces %(Nfaces)d
#define p_Nafp (p_Nfaces*p_Nfp)
#define BSIZE %(BSIZE)d

__kernel void MaxwellsSurface2d(int K,
  __read_only __global float *g_Hx,
  __read_only __global float *g_Hy,
  __read_only __global float *g_Ez,
  __global float *g_rhsHx,
  __global float *g_rhsHy,
  __global float *g_rhsEz,
  read_only __global float *g_surfinfo,
  __read_only image2d_t i_LIFT)
{
  const sampler_t samp =
    CLK_NORMALIZED_COORDS_FALSE
    | CLK_ADDRESS_CLAMP
    | CLK_FILTER_NEAREST;

  /* LOCKED IN to using Np threads per block */
// start_surf_kernel_flux
  const int n = get_local_id(0);
  const int k = get_group_id(0);

  __local float l_fluxHx[p_Nafp];
  __local float l_fluxHy[p_Nafp];
  __local float l_fluxEz[p_Nafp];

  int m;

  /* grab surface nodes and store flux in shared memory */
  if (n < p_Nafp)
  {
    /* coalesced reads (maybe) */
    m = 6*(k*p_Nafp)+n;

    const  int   idM = g_surfinfo[m]; m += p_Nafp;
    int          idP = g_surfinfo[m]; m += p_Nafp;
    const  float Fsc = g_surfinfo[m]; m += p_Nafp;
    const  float Bsc = g_surfinfo[m]; m += p_Nafp;
    const  float nx  = g_surfinfo[m]; m += p_Nafp;
    const  float ny  = g_surfinfo[m];

    float dHx=0, dHy=0, dEz=0;
    dHx = 0.5f*Fsc*(    g_Hx[idP] - g_Hx[idM]);
    dHy = 0.5f*Fsc*(    g_Hy[idP] - g_Hy[idM]);
    dEz = 0.5f*Fsc*(Bsc*g_Ez[idP] - g_Ez[idM]);

    const float ndotdH = nx*dHx + ny*dHy;

    l_fluxHx[n] = -ny*dEz + dHx - ndotdH*nx;
    l_fluxHy[n] =  nx*dEz + dHy - ndotdH*ny;
    l_fluxEz[n] =  nx*dHy - ny*dHx + dEz;
  }

  /* make sure all element data points are cached */
  barrier(CLK_LOCAL_MEM_FENCE);
// end

// start_surf_kernel_lift
  if (n < p_Np)
  {
    float rhsHx = 0, rhsHy = 0, rhsEz = 0;
    int col = 0;

    /* can manually unroll to 3 because there are 3 faces */
    for (m=0;m < p_Nfaces*p_Nfp;)
    {
      float4 L = read_imagef(i_LIFT, samp, (int2)(col, n));
      ++col;

      rhsHx += L.x*l_fluxHx[m];
      rhsHy += L.x*l_fluxHy[m];
      rhsEz += L.x*l_fluxEz[m];
      ++m;

      rhsHx += L.y*l_fluxHx[m];
      rhsHy += L.y*l_fluxHy[m];
      rhsEz += L.y*l_fluxEz[m];
      ++m;

      rhsHx += L.z*l_fluxHx[m];
      rhsHy += L.z*l_fluxHy[m];
      rhsEz += L.z*l_fluxEz[m];
      ++m;
    }

    m = n+k*BSIZE;

    g_rhsHx[m] += rhsHx;
    g_rhsHy[m] += rhsHy;
    g_rhsEz[m] += rhsEz;
  }
// end
}
"""

# }}}




class CLMaxwellsRhs2D:
    def __init__(self, discr):
        import pyopencl as cl
        from pydgeon.opencl import CL_OPTIONS

        self.discr = discr

        self.volume_kernel = cl.Program(discr.ctx,
                MAXWELL2D_VOLUME_KERNEL % {
                    "N": discr.ldis.N,
                    "Np": discr.ldis.Np,
                    "BSIZE": discr.block_size,
                    }
                ).build(options=CL_OPTIONS).MaxwellsVolume2d
        self.volume_kernel.set_scalar_arg_dtypes([np.int32] + 8*[None])

        self.surface_kernel = cl.Program(discr.ctx,
                MAXWELL2D_SURFACE_KERNEL % {
                    "N": discr.ldis.N,
                    "Np": discr.ldis.Np,
                    "Nfp": discr.ldis.Nfp,
                    "Nfaces": discr.ldis.Nfaces,
                    "BSIZE": discr.block_size,
                    }
                ).build(options=CL_OPTIONS).MaxwellsSurface2d
        self.surface_kernel.set_scalar_arg_dtypes([np.int32] + 8*[None])

        self.flops = self.rhs_flops()

    def __call__(self, Hx, Hy, Ez):
        d = self.discr
        ldis = d.ldis

        rhsHx = d.volume_empty()
        rhsHy = d.volume_empty()
        rhsEz = d.volume_empty()

        vol_evt = self.volume_kernel(d.queue, 
                (d.K*d.block_size,), (d.block_size,),
                d.K,
                Hx.data, Hy.data, Ez.data,
                rhsHx.data, rhsHy.data, rhsEz.data,
                d.diffmatrices_img, d.drdx_dev.data)

        surf_block_size = max(
                ldis.Nfp*ldis.Nfaces,
                ldis.Np)

        sfc_evt = self.surface_kernel(d.queue, 
                (d.K*surf_block_size,), (surf_block_size,),
                d.K,
                Hx.data, Hy.data, Ez.data,
                rhsHx.data, rhsHy.data, rhsEz.data,
                d.surfinfo_dev.data, d.lift_img)

        if d.profile >= 5:
            sfc_evt.wait()
            vol_t = (vol_evt.profile.end - vol_evt.profile.start)*1e-9
            sfc_t = (sfc_evt.profile.end - sfc_evt.profile.start)*1e-9
            print "vol: %g s sfc: %g s" % (vol_t, sfc_t)

        return rhsHx, rhsHy, rhsEz

    def rhs_flops(self):
        discr = self.discr
        ldis = discr.ldis
        K = discr.K
        Np = ldis.Np
        Nafp = ldis.Nafp
        d = discr.dimensions

        rs_diff_flops = 2 * K * Np**2
        l2g_diff_flops = 2 * K * Np * (3*4+1)
        lift_flops = K * (2* Np * Nafp + Np) # including jacobian mult
        flux_flops = K * Nafp * (
            3 # jumps
            + 15 # actual upwind flux
            )

        return (6*rs_diff_flops  # each component in r and s
            + l2g_diff_flops 
            + 3*lift_flops # each component
            + flux_flops)

# }}}

# vim: foldmethod=marker
\end{lstlisting}
\caption{
  OpenCL kernel implementing element-local volume contribution
  in the discontinuous Galerkin method, consisting of element-local
  polynomial differentiation.
}
\label{lst:pydgeon-volume-kernel}
\end{proglisting}

One key objective of this subroutine is the multiplication of the
local differentiation matrices $\mathcal D^{\partial \nu}$ by a large number of
right hand sides, each representing degrees of freedom (``DOFs'') on an element.
This is a suitable point to realize that matrix-vector multiplication
by a large number of vectors is equivalent to matrix-matrix
multiplication by a very fat, moderately short matrix that encompasses
all elemental vectors glued together. Perhaps the most immediate
approach to such a problem might be to use Nvidia's CUBLAS.
Unfortunately, while CUBLAS successfully covers a great many use
cases, the matrix sizes in question here resulted in uninspiring
performance in our experiments \citep{kloeckner_nodal_2009}. We are
thus left considering the choices for a from-scratch implementation.

In the design of computational kernels for GPUs, perhaps \emph{the}
key defining factor is the work decomposition into thread blocks (in
CUDA terminology) or work groups (in OpenCL terminology). In our
demonstration solver, we choose a very simple alternative, a
one-to-one mapping between elements and work groups, and a one-to-one
mapping between output degrees of freedom and threads (CUDA) or work
items (OpenCL).  This choice is simple and expedient, but it can be
improved upon in a number of cases, as we will discuss in Section
\ref{sec:work-partitition}.

The next key decision is the memory layout of the data to be worked
on. Again, we make a simple choice and describe possible
improvements later. As was discussed in Section
\ref{sec:dg-method-construction}, the data we are working on consists
of $N_p$ coefficients of Lagrange interpolation polynomials for each
of the $K$ elements. Observe that, by their being coefficients of
interpolation polynomials, they each represent the exact value of the
represented solution at a point in space belonging to a certain
element. 

To ensure that each work group can fetch element data in the least
number of memory transactions, we choose to pad each element up to
$\lceil N_p\rceil_{16}$ floating point values, where the notation
$\lceil x \rceil_{y}$ represents $x$ rounded up to the nearest
multiple of $y$.

With data layout and work decomposition clarified,we can now examine
the implementation itself, as shown in Listing
\ref{lst:pydgeon-volume-kernel}.  After getting the element number
\textsf{k} and the number of the elemental degree of freedom
\textsf{n} from group and local IDs, respectively, first the elemental
degrees of freedom for all three fields ($H_x$, $H_y$, $E_z$) are
fetched into local memory for subsequent multiplication by the
differentiation matrix. 

As indicated above, the work being performed is effectively
matrix-matrix multiplication, and therefore existing best practices
suggest that also fetching the matrix into local memory might be a
good idea. At least on pre-Fermi chips, this does not turn out to be
true. We will take a closer look at the trade-offs involved in Section
\ref{sec:which-memory-for-what}. For now, we simply state that the
matrix is streamed into core through texture memory, and its fetch
cost amortized by reusing it for not just one, but all three fields
($H_x$, $H_y$, and $E_z$).

Once the derivatives along each element's axes are computed by matrix
multiplication, they are converted to global $x$ and $y$ derivatives
according to \eqref{eq:dg-diff-local-to-global}, using separate
per-element geometric factors. Finally, the results are stored, where
our memory layout and work decomposition permit a fully coalesced
write.

\subsubsection{Computing the Surface Contribution}
\label{sec:dg-surface-kernel}

\begin{proglisting}
\begin{lstlisting}[linerange=start_surf_kernel_flux-end]
# Pydgeon - the Python DG Environment
# (C) 2009, 2010 Tim Warburton, Xueyu Zhu, Andreas Kloeckner
#
# This program is free software: you can redistribute it and/or modify
# it under the terms of the GNU General Public License as published by
# the Free Software Foundation, either version 3 of the License, or
# (at your option) any later version.
#
# This program is distributed in the hope that it will be useful,
# but WITHOUT ANY WARRANTY; without even the implied warranty of
# MERCHANTABILITY or FITNESS FOR A PARTICULAR PURPOSE.  See the
# GNU General Public License for more details.
#
# You should have received a copy of the GNU General Public License
# along with this program.  If not, see <http://www.gnu.org/licenses/>.



from __future__ import division

import numpy as np




# {{{ CPU

def MaxwellRHS2D(discr, Hx, Hy, Ez):
    """Evaluate RHS flux in 2D Maxwell TM form."""

    from pydgeon.tools import eldot

    d = discr
    l = discr.ldis

    # Define field differences at faces
    vmapM = d.vmapM.reshape(d.K, -1)
    vmapP = d.vmapP.reshape(d.K, -1)

    dHx = Hx.flat[vmapM]-Hx.flat[vmapP]
    dHy = Hy.flat[vmapM]-Hy.flat[vmapP]
    dEz = Ez.flat[vmapM]-Ez.flat[vmapP]

    # Impose reflective boundary conditions (Ez+ = -Ez-)
    dHx.flat[d.mapB] = 0
    dHy.flat[d.mapB] = 0
    dEz.flat[d.mapB] = 2*Ez.flat[d.vmapB]

    # evaluate upwind fluxes
    alpha  = 1.0
    ndotdH =  d.nx*dHx + d.ny*dHy
    fluxHx =  d.ny*dEz + alpha*(ndotdH*d.nx-dHx)
    fluxHy = -d.nx*dEz + alpha*(ndotdH*d.ny-dHy)
    fluxEz = -d.nx*dHy + d.ny*dHx - alpha*dEz

    # local derivatives of fields
    Ezx, Ezy = d.grad(Ez)
    CuHx, CuHy, CuHz = d.curl(Hx, Hy,0)

    # compute right hand sides of the PDE's
    rhsHx = -Ezy  + eldot(l.LIFT, (d.Fscale*fluxHx))/2.0
    rhsHy =  Ezx  + eldot(l.LIFT, (d.Fscale*fluxHy))/2.0
    rhsEz =  CuHz + eldot(l.LIFT, (d.Fscale*fluxEz))/2.0
    return rhsHx, rhsHy, rhsEz

# }}}

# {{{ OpenCL

# {{{ kernels

MAXWELL2D_VOLUME_KERNEL = """
#define p_N %(N)d
#define p_Np %(Np)d
#define BSIZE %(BSIZE)d

__kernel void MaxwellsVolume2d(int K,
   __read_only __global float *g_Hx,
   __read_only __global float *g_Hy,
   __read_only __global float *g_Ez,
   __global float *g_rhsHx,
   __global float *g_rhsHy,
   __global float *g_rhsEz,
   __read_only __global image2d_t i_DrDs,
   __read_only __global float *g_vgeo)
{
  const sampler_t samp =
    CLK_NORMALIZED_COORDS_FALSE
    | CLK_ADDRESS_CLAMP
    | CLK_FILTER_NEAREST;

  __local float l_Hx[BSIZE];
  __local float l_Hy[BSIZE];
  __local float l_Ez[BSIZE];

  /* LOCKED IN to using Np work items per group */
// start_vol_kernel
  const int n = get_local_id(0);
  const int k = get_group_id(0);

  int m = n+k*BSIZE;
  int id = n;

  l_Hx[id] = g_Hx[m];
  l_Hy[id] = g_Hy[m];
  l_Ez[id] = g_Ez[m];

  barrier(CLK_LOCAL_MEM_FENCE);

  float dHxdr=0,dHxds=0;
  float dHydr=0,dHyds=0;
  float dEzdr=0,dEzds=0;

  float Q;
  for(m=0; m<p_Np; ++m)
  {
    float4 D = read_imagef(i_DrDs, samp, (int2)(n, m));

    Q = l_Hx[m]; dHxdr += D.x*Q; dHxds += D.y*Q;
    Q = l_Hy[m]; dHydr += D.x*Q; dHyds += D.y*Q;
    Q = l_Ez[m]; dEzdr += D.x*Q; dEzds += D.y*Q;
  }

  const float drdx = g_vgeo[0+4*k];
  const float drdy = g_vgeo[1+4*k];
  const float dsdx = g_vgeo[2+4*k];
  const float dsdy = g_vgeo[3+4*k];

  m = n+BSIZE*k;
  g_rhsHx[m] = -(drdy*dEzdr+dsdy*dEzds);
  g_rhsHy[m] =  (drdx*dEzdr+dsdx*dEzds);
  g_rhsEz[m] =  (drdx*dHydr+dsdx*dHyds 
    - drdy*dHxdr-dsdy*dHxds);
// end
}
"""

MAXWELL2D_SURFACE_KERNEL = """
#define p_N %(N)d
#define p_Np %(Np)d
#define p_Nfp %(Nfp)d
#define p_Nfaces %(Nfaces)d
#define p_Nafp (p_Nfaces*p_Nfp)
#define BSIZE %(BSIZE)d

__kernel void MaxwellsSurface2d(int K,
  __read_only __global float *g_Hx,
  __read_only __global float *g_Hy,
  __read_only __global float *g_Ez,
  __global float *g_rhsHx,
  __global float *g_rhsHy,
  __global float *g_rhsEz,
  read_only __global float *g_surfinfo,
  __read_only image2d_t i_LIFT)
{
  const sampler_t samp =
    CLK_NORMALIZED_COORDS_FALSE
    | CLK_ADDRESS_CLAMP
    | CLK_FILTER_NEAREST;

  /* LOCKED IN to using Np threads per block */
// start_surf_kernel_flux
  const int n = get_local_id(0);
  const int k = get_group_id(0);

  __local float l_fluxHx[p_Nafp];
  __local float l_fluxHy[p_Nafp];
  __local float l_fluxEz[p_Nafp];

  int m;

  /* grab surface nodes and store flux in shared memory */
  if (n < p_Nafp)
  {
    /* coalesced reads (maybe) */
    m = 6*(k*p_Nafp)+n;

    const  int   idM = g_surfinfo[m]; m += p_Nafp;
    int          idP = g_surfinfo[m]; m += p_Nafp;
    const  float Fsc = g_surfinfo[m]; m += p_Nafp;
    const  float Bsc = g_surfinfo[m]; m += p_Nafp;
    const  float nx  = g_surfinfo[m]; m += p_Nafp;
    const  float ny  = g_surfinfo[m];

    float dHx=0, dHy=0, dEz=0;
    dHx = 0.5f*Fsc*(    g_Hx[idP] - g_Hx[idM]);
    dHy = 0.5f*Fsc*(    g_Hy[idP] - g_Hy[idM]);
    dEz = 0.5f*Fsc*(Bsc*g_Ez[idP] - g_Ez[idM]);

    const float ndotdH = nx*dHx + ny*dHy;

    l_fluxHx[n] = -ny*dEz + dHx - ndotdH*nx;
    l_fluxHy[n] =  nx*dEz + dHy - ndotdH*ny;
    l_fluxEz[n] =  nx*dHy - ny*dHx + dEz;
  }

  /* make sure all element data points are cached */
  barrier(CLK_LOCAL_MEM_FENCE);
// end

// start_surf_kernel_lift
  if (n < p_Np)
  {
    float rhsHx = 0, rhsHy = 0, rhsEz = 0;
    int col = 0;

    /* can manually unroll to 3 because there are 3 faces */
    for (m=0;m < p_Nfaces*p_Nfp;)
    {
      float4 L = read_imagef(i_LIFT, samp, (int2)(col, n));
      ++col;

      rhsHx += L.x*l_fluxHx[m];
      rhsHy += L.x*l_fluxHy[m];
      rhsEz += L.x*l_fluxEz[m];
      ++m;

      rhsHx += L.y*l_fluxHx[m];
      rhsHy += L.y*l_fluxHy[m];
      rhsEz += L.y*l_fluxEz[m];
      ++m;

      rhsHx += L.z*l_fluxHx[m];
      rhsHy += L.z*l_fluxHy[m];
      rhsEz += L.z*l_fluxEz[m];
      ++m;
    }

    m = n+k*BSIZE;

    g_rhsHx[m] += rhsHx;
    g_rhsHy[m] += rhsHy;
    g_rhsEz[m] += rhsEz;
  }
// end
}
"""

# }}}




class CLMaxwellsRhs2D:
    def __init__(self, discr):
        import pyopencl as cl
        from pydgeon.opencl import CL_OPTIONS

        self.discr = discr

        self.volume_kernel = cl.Program(discr.ctx,
                MAXWELL2D_VOLUME_KERNEL % {
                    "N": discr.ldis.N,
                    "Np": discr.ldis.Np,
                    "BSIZE": discr.block_size,
                    }
                ).build(options=CL_OPTIONS).MaxwellsVolume2d
        self.volume_kernel.set_scalar_arg_dtypes([np.int32] + 8*[None])

        self.surface_kernel = cl.Program(discr.ctx,
                MAXWELL2D_SURFACE_KERNEL % {
                    "N": discr.ldis.N,
                    "Np": discr.ldis.Np,
                    "Nfp": discr.ldis.Nfp,
                    "Nfaces": discr.ldis.Nfaces,
                    "BSIZE": discr.block_size,
                    }
                ).build(options=CL_OPTIONS).MaxwellsSurface2d
        self.surface_kernel.set_scalar_arg_dtypes([np.int32] + 8*[None])

        self.flops = self.rhs_flops()

    def __call__(self, Hx, Hy, Ez):
        d = self.discr
        ldis = d.ldis

        rhsHx = d.volume_empty()
        rhsHy = d.volume_empty()
        rhsEz = d.volume_empty()

        vol_evt = self.volume_kernel(d.queue, 
                (d.K*d.block_size,), (d.block_size,),
                d.K,
                Hx.data, Hy.data, Ez.data,
                rhsHx.data, rhsHy.data, rhsEz.data,
                d.diffmatrices_img, d.drdx_dev.data)

        surf_block_size = max(
                ldis.Nfp*ldis.Nfaces,
                ldis.Np)

        sfc_evt = self.surface_kernel(d.queue, 
                (d.K*surf_block_size,), (surf_block_size,),
                d.K,
                Hx.data, Hy.data, Ez.data,
                rhsHx.data, rhsHy.data, rhsEz.data,
                d.surfinfo_dev.data, d.lift_img)

        if d.profile >= 5:
            sfc_evt.wait()
            vol_t = (vol_evt.profile.end - vol_evt.profile.start)*1e-9
            sfc_t = (sfc_evt.profile.end - sfc_evt.profile.start)*1e-9
            print "vol: %g s sfc: %g s" % (vol_t, sfc_t)

        return rhsHx, rhsHy, rhsEz

    def rhs_flops(self):
        discr = self.discr
        ldis = discr.ldis
        K = discr.K
        Np = ldis.Np
        Nafp = ldis.Nafp
        d = discr.dimensions

        rs_diff_flops = 2 * K * Np**2
        l2g_diff_flops = 2 * K * Np * (3*4+1)
        lift_flops = K * (2* Np * Nafp + Np) # including jacobian mult
        flux_flops = K * Nafp * (
            3 # jumps
            + 15 # actual upwind flux
            )

        return (6*rs_diff_flops  # each component in r and s
            + l2g_diff_flops 
            + 3*lift_flops # each component
            + flux_flops)

# }}}

# vim: foldmethod=marker
\end{lstlisting}
(Continued in Listing \ref{lst:pydgeon-surface-kernel-lift}.)
\caption{
  Part 1 of the OpenCL kernel implementing inter-element surface 
  contribution in the discontinuous Galerkin method, consisting 
  of the calculation of the surface flux of \eqref{eq:max2d-numflux}.
}
\label{lst:pydgeon-surface-kernel-flux}
\end{proglisting}

\begin{proglisting}
(Continued from Figure \ref{lst:pydgeon-surface-kernel-flux}.)
\begin{lstlisting}[linerange=start_surf_kernel_lift-end]
# Pydgeon - the Python DG Environment
# (C) 2009, 2010 Tim Warburton, Xueyu Zhu, Andreas Kloeckner
#
# This program is free software: you can redistribute it and/or modify
# it under the terms of the GNU General Public License as published by
# the Free Software Foundation, either version 3 of the License, or
# (at your option) any later version.
#
# This program is distributed in the hope that it will be useful,
# but WITHOUT ANY WARRANTY; without even the implied warranty of
# MERCHANTABILITY or FITNESS FOR A PARTICULAR PURPOSE.  See the
# GNU General Public License for more details.
#
# You should have received a copy of the GNU General Public License
# along with this program.  If not, see <http://www.gnu.org/licenses/>.



from __future__ import division

import numpy as np




# {{{ CPU

def MaxwellRHS2D(discr, Hx, Hy, Ez):
    """Evaluate RHS flux in 2D Maxwell TM form."""

    from pydgeon.tools import eldot

    d = discr
    l = discr.ldis

    # Define field differences at faces
    vmapM = d.vmapM.reshape(d.K, -1)
    vmapP = d.vmapP.reshape(d.K, -1)

    dHx = Hx.flat[vmapM]-Hx.flat[vmapP]
    dHy = Hy.flat[vmapM]-Hy.flat[vmapP]
    dEz = Ez.flat[vmapM]-Ez.flat[vmapP]

    # Impose reflective boundary conditions (Ez+ = -Ez-)
    dHx.flat[d.mapB] = 0
    dHy.flat[d.mapB] = 0
    dEz.flat[d.mapB] = 2*Ez.flat[d.vmapB]

    # evaluate upwind fluxes
    alpha  = 1.0
    ndotdH =  d.nx*dHx + d.ny*dHy
    fluxHx =  d.ny*dEz + alpha*(ndotdH*d.nx-dHx)
    fluxHy = -d.nx*dEz + alpha*(ndotdH*d.ny-dHy)
    fluxEz = -d.nx*dHy + d.ny*dHx - alpha*dEz

    # local derivatives of fields
    Ezx, Ezy = d.grad(Ez)
    CuHx, CuHy, CuHz = d.curl(Hx, Hy,0)

    # compute right hand sides of the PDE's
    rhsHx = -Ezy  + eldot(l.LIFT, (d.Fscale*fluxHx))/2.0
    rhsHy =  Ezx  + eldot(l.LIFT, (d.Fscale*fluxHy))/2.0
    rhsEz =  CuHz + eldot(l.LIFT, (d.Fscale*fluxEz))/2.0
    return rhsHx, rhsHy, rhsEz

# }}}

# {{{ OpenCL

# {{{ kernels

MAXWELL2D_VOLUME_KERNEL = """
#define p_N %(N)d
#define p_Np %(Np)d
#define BSIZE %(BSIZE)d

__kernel void MaxwellsVolume2d(int K,
   __read_only __global float *g_Hx,
   __read_only __global float *g_Hy,
   __read_only __global float *g_Ez,
   __global float *g_rhsHx,
   __global float *g_rhsHy,
   __global float *g_rhsEz,
   __read_only __global image2d_t i_DrDs,
   __read_only __global float *g_vgeo)
{
  const sampler_t samp =
    CLK_NORMALIZED_COORDS_FALSE
    | CLK_ADDRESS_CLAMP
    | CLK_FILTER_NEAREST;

  __local float l_Hx[BSIZE];
  __local float l_Hy[BSIZE];
  __local float l_Ez[BSIZE];

  /* LOCKED IN to using Np work items per group */
// start_vol_kernel
  const int n = get_local_id(0);
  const int k = get_group_id(0);

  int m = n+k*BSIZE;
  int id = n;

  l_Hx[id] = g_Hx[m];
  l_Hy[id] = g_Hy[m];
  l_Ez[id] = g_Ez[m];

  barrier(CLK_LOCAL_MEM_FENCE);

  float dHxdr=0,dHxds=0;
  float dHydr=0,dHyds=0;
  float dEzdr=0,dEzds=0;

  float Q;
  for(m=0; m<p_Np; ++m)
  {
    float4 D = read_imagef(i_DrDs, samp, (int2)(n, m));

    Q = l_Hx[m]; dHxdr += D.x*Q; dHxds += D.y*Q;
    Q = l_Hy[m]; dHydr += D.x*Q; dHyds += D.y*Q;
    Q = l_Ez[m]; dEzdr += D.x*Q; dEzds += D.y*Q;
  }

  const float drdx = g_vgeo[0+4*k];
  const float drdy = g_vgeo[1+4*k];
  const float dsdx = g_vgeo[2+4*k];
  const float dsdy = g_vgeo[3+4*k];

  m = n+BSIZE*k;
  g_rhsHx[m] = -(drdy*dEzdr+dsdy*dEzds);
  g_rhsHy[m] =  (drdx*dEzdr+dsdx*dEzds);
  g_rhsEz[m] =  (drdx*dHydr+dsdx*dHyds 
    - drdy*dHxdr-dsdy*dHxds);
// end
}
"""

MAXWELL2D_SURFACE_KERNEL = """
#define p_N %(N)d
#define p_Np %(Np)d
#define p_Nfp %(Nfp)d
#define p_Nfaces %(Nfaces)d
#define p_Nafp (p_Nfaces*p_Nfp)
#define BSIZE %(BSIZE)d

__kernel void MaxwellsSurface2d(int K,
  __read_only __global float *g_Hx,
  __read_only __global float *g_Hy,
  __read_only __global float *g_Ez,
  __global float *g_rhsHx,
  __global float *g_rhsHy,
  __global float *g_rhsEz,
  read_only __global float *g_surfinfo,
  __read_only image2d_t i_LIFT)
{
  const sampler_t samp =
    CLK_NORMALIZED_COORDS_FALSE
    | CLK_ADDRESS_CLAMP
    | CLK_FILTER_NEAREST;

  /* LOCKED IN to using Np threads per block */
// start_surf_kernel_flux
  const int n = get_local_id(0);
  const int k = get_group_id(0);

  __local float l_fluxHx[p_Nafp];
  __local float l_fluxHy[p_Nafp];
  __local float l_fluxEz[p_Nafp];

  int m;

  /* grab surface nodes and store flux in shared memory */
  if (n < p_Nafp)
  {
    /* coalesced reads (maybe) */
    m = 6*(k*p_Nafp)+n;

    const  int   idM = g_surfinfo[m]; m += p_Nafp;
    int          idP = g_surfinfo[m]; m += p_Nafp;
    const  float Fsc = g_surfinfo[m]; m += p_Nafp;
    const  float Bsc = g_surfinfo[m]; m += p_Nafp;
    const  float nx  = g_surfinfo[m]; m += p_Nafp;
    const  float ny  = g_surfinfo[m];

    float dHx=0, dHy=0, dEz=0;
    dHx = 0.5f*Fsc*(    g_Hx[idP] - g_Hx[idM]);
    dHy = 0.5f*Fsc*(    g_Hy[idP] - g_Hy[idM]);
    dEz = 0.5f*Fsc*(Bsc*g_Ez[idP] - g_Ez[idM]);

    const float ndotdH = nx*dHx + ny*dHy;

    l_fluxHx[n] = -ny*dEz + dHx - ndotdH*nx;
    l_fluxHy[n] =  nx*dEz + dHy - ndotdH*ny;
    l_fluxEz[n] =  nx*dHy - ny*dHx + dEz;
  }

  /* make sure all element data points are cached */
  barrier(CLK_LOCAL_MEM_FENCE);
// end

// start_surf_kernel_lift
  if (n < p_Np)
  {
    float rhsHx = 0, rhsHy = 0, rhsEz = 0;
    int col = 0;

    /* can manually unroll to 3 because there are 3 faces */
    for (m=0;m < p_Nfaces*p_Nfp;)
    {
      float4 L = read_imagef(i_LIFT, samp, (int2)(col, n));
      ++col;

      rhsHx += L.x*l_fluxHx[m];
      rhsHy += L.x*l_fluxHy[m];
      rhsEz += L.x*l_fluxEz[m];
      ++m;

      rhsHx += L.y*l_fluxHx[m];
      rhsHy += L.y*l_fluxHy[m];
      rhsEz += L.y*l_fluxEz[m];
      ++m;

      rhsHx += L.z*l_fluxHx[m];
      rhsHy += L.z*l_fluxHy[m];
      rhsEz += L.z*l_fluxEz[m];
      ++m;
    }

    m = n+k*BSIZE;

    g_rhsHx[m] += rhsHx;
    g_rhsHy[m] += rhsHy;
    g_rhsEz[m] += rhsEz;
  }
// end
}
"""

# }}}




class CLMaxwellsRhs2D:
    def __init__(self, discr):
        import pyopencl as cl
        from pydgeon.opencl import CL_OPTIONS

        self.discr = discr

        self.volume_kernel = cl.Program(discr.ctx,
                MAXWELL2D_VOLUME_KERNEL % {
                    "N": discr.ldis.N,
                    "Np": discr.ldis.Np,
                    "BSIZE": discr.block_size,
                    }
                ).build(options=CL_OPTIONS).MaxwellsVolume2d
        self.volume_kernel.set_scalar_arg_dtypes([np.int32] + 8*[None])

        self.surface_kernel = cl.Program(discr.ctx,
                MAXWELL2D_SURFACE_KERNEL % {
                    "N": discr.ldis.N,
                    "Np": discr.ldis.Np,
                    "Nfp": discr.ldis.Nfp,
                    "Nfaces": discr.ldis.Nfaces,
                    "BSIZE": discr.block_size,
                    }
                ).build(options=CL_OPTIONS).MaxwellsSurface2d
        self.surface_kernel.set_scalar_arg_dtypes([np.int32] + 8*[None])

        self.flops = self.rhs_flops()

    def __call__(self, Hx, Hy, Ez):
        d = self.discr
        ldis = d.ldis

        rhsHx = d.volume_empty()
        rhsHy = d.volume_empty()
        rhsEz = d.volume_empty()

        vol_evt = self.volume_kernel(d.queue, 
                (d.K*d.block_size,), (d.block_size,),
                d.K,
                Hx.data, Hy.data, Ez.data,
                rhsHx.data, rhsHy.data, rhsEz.data,
                d.diffmatrices_img, d.drdx_dev.data)

        surf_block_size = max(
                ldis.Nfp*ldis.Nfaces,
                ldis.Np)

        sfc_evt = self.surface_kernel(d.queue, 
                (d.K*surf_block_size,), (surf_block_size,),
                d.K,
                Hx.data, Hy.data, Ez.data,
                rhsHx.data, rhsHy.data, rhsEz.data,
                d.surfinfo_dev.data, d.lift_img)

        if d.profile >= 5:
            sfc_evt.wait()
            vol_t = (vol_evt.profile.end - vol_evt.profile.start)*1e-9
            sfc_t = (sfc_evt.profile.end - sfc_evt.profile.start)*1e-9
            print "vol: %g s sfc: %g s" % (vol_t, sfc_t)

        return rhsHx, rhsHy, rhsEz

    def rhs_flops(self):
        discr = self.discr
        ldis = discr.ldis
        K = discr.K
        Np = ldis.Np
        Nafp = ldis.Nafp
        d = discr.dimensions

        rs_diff_flops = 2 * K * Np**2
        l2g_diff_flops = 2 * K * Np * (3*4+1)
        lift_flops = K * (2* Np * Nafp + Np) # including jacobian mult
        flux_flops = K * Nafp * (
            3 # jumps
            + 15 # actual upwind flux
            )

        return (6*rs_diff_flops  # each component in r and s
            + l2g_diff_flops 
            + 3*lift_flops # each component
            + flux_flops)

# }}}

# vim: foldmethod=marker
\end{lstlisting}
\caption{
  Part 2 of the OpenCL kernel implementing inter-element surface 
  contribution in the discontinuous Galerkin method, consisting 
  of the lifting of the surface flux contribution, as described
  in \eqref{eq:dg-lift}. 
}
\label{lst:pydgeon-surface-kernel-lift}
\end{proglisting}

The second (and slightly more complicated) part of our sample
implementation of the DG method is what we call the ``surface
kernel'', which, as part of the same subroutine, achieves both the
extraction of the flux expression of \eqref{eq:max2d-numflux} and the
surface integration of \eqref{eq:dg-lift}. The key parts of the OpenCL
C source code of the kernel is shown in
Listing~\ref{lst:pydgeon-surface-kernel-flux} and continued in
Listing~\ref{lst:pydgeon-surface-kernel-lift}.

We use the same one-work-group-per-element work partition as in the
previous section, and obviously the memory layout of the element data
is likewise unchanged. It is however important to note that the kernel
in question here operates on two different data formats during its
lifetime. First, the output of the flux gather results in a vector
of facial degrees of freedom as displayed in \eqref{eq:dg-lift}.
The number of entries in this vector is $3N_{fp}$, where three is the
number of faces in a triangle, and $N_{fp}=N+1$ is the number of
degrees of freedom required to discretize each face. 
In general $3N_{fp}\ne N_p$, where we recall that $N_p$ is the number
of volume degrees of freedom. At each of the two stages of the
algorithm, we employ a design that uses one work item per degree of
freedom. The number of work items required per work group is therefore
$\max(N_p, N_{fp})$, and at the start of each stage of the algorithm,
we need to verify whether the local thread number is less than the
number of outputs required in that stage, to avoid computing (and
perhaps storing) spurious extra outputs. This is necessary in both
stages because either of $N_p$ or $3N_{fp}$ may be larger.

After fixing the DOF and element indices \textsf{n} and \textsf{k},
the kernel begins by allocating $3N_{fp}$ degrees of freedom of local
storage ($N_{fp}$ per face) for each of the three fields ($H_x$, $H_y$, and $E_z$). This
local memory serves as a temporary storage for the facial vector of
\eqref{eq:dg-lift}. Next, index and geometry information is read from
a surface descriptor data structure called \textsf{surfinfo}.  For
each facial degree of freedom, and hence for each work item, this data
structure contains the index of the volume degree of freedom the work
item processes, as well as the index of its facial neighbor
(\textsf{idM} and \textsf{idP}, respectively). In addition,
\textsf{surfinfo} contains geometry information, namely the surface
unit normal of the face being integrated over (\textsf{nx} and
\textsf{ny}), the surface Jacobian divided by the element's volume
Jacobian (\textsf{Fsc}) and a boundary indicator (\textsf{Bsc}) used
for the implementation of boundary conditions which takes values of
$\pm 1$. Based on this information, the kernel computes jump terms
$[H_x]$, $[H_y]$, $[E_z]$ which are then scaled with the geometry
scaling \textsf{Fsc} and stored as \textsf{dHx}, \textsf{dHy}, and
\textsf{dEz}.  The computation of the flux expression
\eqref{eq:max2d-numflux} and its temporary storage in local memory
concludes the first part of the kernel, displayed in Listing
\ref{lst:pydgeon-surface-kernel-flux}.

The second part of the kernel, of Listing
\ref{lst:pydgeon-surface-kernel-lift}, is much like local polynomial
differentiation as discussed in Section \ref{sec:dg-volume-kernel} in
that it represents an element-local matrix multiplication. We have
applied the same design decisions as above for simplicity, mainly
based on the facial flux data already being resident in local memory.
Again, data for the matrix is streamed in using the texture units, and
the streamed matrix is reused for each of the three fields. One
trick we were able to apply here is the three-fold unrolling of the
loop. This is valid because we know that the combined face mass matrix
$\mathcal M^{\partial \I}$ covers three faces and hence must have a
column count that is divisible by three. Naturally, the same applies
to the lifting matrix $\mathcal L := \mathcal M^{-1} \mathcal
M^{\partial \I}$.  The result of this matrix-vector product is then
added to global destination arrays in which the volume contribution to
the right-hand side $\partial (H_x,H_y,E_z)^T/\partial t$ is already
stored, completing the computation the entire right-hand side of
\eqref{eq:dg-strong-explicit}.

This concludes our description of the basic kernel implementing the
computation of the ODE right-hand side for nodal discontinuous
Galerkin methods. What is missing to complete the implementation
of the method is a simple Runge-Kutta time integrator, which we have
implemented using PyOpenCL's built-in array operations. We now proceed
to discuss a number of ways in which performance of these basic
kernels can be improved.

\subsection{Improving Performance}
\label{sec:improving-performance}

\subsubsection{Avoiding Padding Waste: Data Aggregation }
\label{sec:data-aggregation}
In the above codes, each element is represented by $\lceil N_p
\rceil_{16}$ floating point values for alignment and fetch efficiency
reasons. Especially in two dimensions, or in three dimensions for
elements of relatively small polynomial order $N$, this extra padding
can be rather inefficient--not just in terms of GPU memory use, but
especially also in computational resources. All of our kernels adopt
a one-work-item-per-output design, and hence wasted memory has a
one-to-one correspondence to wasted computational power. This is all
the more true once one realizes that Nvidia hardware schedules
computations in units of 32-wide \emph{warps}, such that a rounding to
16 has a chance of 50 per cent of leaving the trailing half-warp of
the computation unused. An obvious remedy for this problem is the
aggregation of multiple elements into a single unit. This aggregation
represents a trade-off against the work partition flexibility of all
kernels operating on the data, and should therefore be chosen as small
as possible, while still minimizing waste.

In \citep{kloeckner_nodal_2009}, we pursue a compromise strategy,
where we choose a granularity that combines enough elements so that less
than a given percentage (e.g. 10) of waste occurs. All further
occurring granularities are then required to operate
integer multiples of this smallest possible granularity.  To
differentiate this granularity from the generally larger work group
size (or ``block size'' in CUDA terminology), we have introduced the
term \emph{microblock} to denote it.
\subsubsection{Which Memory for What?}
\label{sec:which-memory-for-what}
On-chip memory in a GPU setting is always somewhat scarce, and we
foresee that this will remain so for the foreseeable future. As
already discussed in Section \ref{sec:dg-minimal-implementation}, it
is far from clear which portions of the GPU's on-chip memory
should be used for what data. In discussing this question, we focus on
the element-local matrix-vector polynomial differentiation as this
asymptotically (and practically) dominates runtime as the
polynomial degree $N$ increases.

In \citep{kloeckner_nodal_2009}, we discuss two possible strategies
for element-local matrix multiplication, the first of which proceeds
by loading the matrix into local memory, and the second of which loads
field data into local memory, as we have done above. 

To allow the flexibility of being able to choose which strategy to use
for each each of the two element-local matrix products
(differentiation and lift), the surface kernel of Section
\ref{sec:dg-surface-kernel} may have to be split into its two
constituent parts, necessitating an extra store-load cycle. We find
that this disadvantage is entirely compensated by the advantage of
being able to use a more immediately suitable work partition for each
part.

The enumeration of the two strategies begs an immediate question--why
is the strategy of loading \emph{both} quantities into local memory
not considered? The reason for this lies rooted in a number of important
practicalities. While generic matrix multiplication routines are free
to optimize for the case of large, square matrices, the matrices we
are faced with are small. A generic blocking strategy would therefore
leave us with many inefficient corner cases which would come to
dominate our run time. In addition, in the case of three-dimensional
geometry,  the three differentiation matrices of size $N_p\times N_p$
exhaust the local memory on Nvidia hardware even for moderate $N$.

We find that, at low-to-moderate $N$ and in general in two dimensions,
we can derive a gain of about 20 per cent by making the
matrix-in-local strategy available in addition to the field-in-local
strategy shown above. Nonetheless, the latter strategy has fewer size
restrictions than the former, is thus more generally applicable, and
it successfully uses register and texture memory to avoid many
redundant matrix fetches. This justifies our choice of the strategy in
our demonstration code. In these codes, we amortized matrix fetch
costs by operating on three fields at once. Note that even in a scalar
(i.e. single-field) case, such amortization is possible, simply by
operating on multiple elements within the same work item.
\subsubsection{Rethinking the Work Partition}
\label{sec:work-partitition}
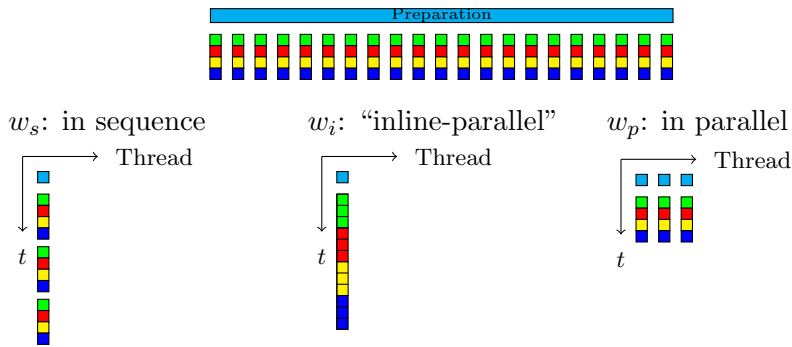
\begin{figure}
  \centering
  \beginpgfgraphicnamed{work-decomposition}
  \begin{minipage}{5in}
  \centering
  \def\workunit#1{
    \draw [draw,fill=green] #1 ++(0,0) rectangle +(0.15,-0.15) ;
    \draw [draw,fill=red] #1 ++(0,-0.15) rectangle +(0.15,-0.15) ;
    \draw [draw,fill=yellow] #1 ++(0,-0.3) rectangle +(0.15,-0.15) ;
    \draw [draw,fill=blue] #1 ++(0,-0.45) rectangle +(0.15,-0.15) ;
  }
  \newcommand\drawaxes{
    \draw [->] (-0.2,0.5) -- +(1,0) node [pos=1,right=0.1cm,font=\footnotesize] {Thread};
    \draw [->] (-0.2,0.5) -- +(0,-1) node [pos=1,below=0.1cm,font=\footnotesize] {$t$};
  }
  \tikzstyle{wuprep}=[draw,fill=cyan]

  \begin{tikzpicture}
    \node at (0,0.15) [anchor=south west,font=\tiny,
      wuprep,minimum width=6.15cm,inner sep=0]
      {Preparation};
    \foreach \x in {0,0.3,...,6.1} { \workunit{(\x,0)}  }
  \end{tikzpicture}

  \vspace{0.3cm}
  \begin{tabular}{p{0.28\textwidth}p{0.28\textwidth}p{0.28\textwidth}}
  $w_s$: in sequence

  \begin{tikzpicture}
    \drawaxes
    \workunit{(0,0)}
    \workunit{(0,-0.7)}
    \workunit{(0,-1.4)}
    \draw [wuprep] (0,0.15) rectangle +(0.15,0.15) ;
  \end{tikzpicture}
  &
  $w_i$: ``inline-parallel''

  \begin{tikzpicture}
    \drawaxes
    \draw [fill=green] (0,0) rectangle +(0.15,-0.45) ;
    \draw [fill=red] (0,-0.45) rectangle +(0.15,-0.45) ;
    \draw [fill=yellow] (0,-0.9) rectangle +(0.15,-0.45) ;
    \draw [fill=blue] (0,-1.35) rectangle +(0.15,-0.45) ;

    \draw [draw,ystep=0.14999cm] (0,-1.8) grid +(0.15,1.8) ;

    \draw [wuprep] (0,0.15) rectangle +(0.15,0.15) ;
  \end{tikzpicture}
  &
  $w_p$: in parallel

  \begin{tikzpicture}
    \drawaxes
    \workunit{(0,0)}
    \workunit{(0.3,0)}
    \workunit{(0.6,0)}
    \draw [wuprep] (0,0.15) rectangle +(0.15,0.15) ;
    \draw [wuprep] (0.3,0.15) rectangle +(0.15,0.15) ;
    \draw [wuprep] (0.6,0.15) rectangle +(0.15,0.15) ;
  \end{tikzpicture}
  \end{tabular}
  \end{minipage}
  \endpgfgraphicnamed
  \caption{Possibilities for partitioning a large
    number independent work units, potentially requiring some
    preparation, into work groups. Each work unit consists
    of multiple stages, symbolized by different colors.
    Preparation is shown in a cyan color. An example of this 
    would be multiple independent matrix-vector multiplications, 
    each consisting of individual multiply-add cycles.
    }
  \label{fig:work-partition}
\end{figure}
Because of the inherent advantages of using one work item per output
value, the question of work partitioning into work groups and work
items on GPU hardware is never far removed from that of memory use and
data layout, as discussed above. The work partition chosen in the
demonstration code was purposefully simple--one work item per degree
of freedom, one work group per element. Moving beyond that, while
taking into account the lessons of Section \ref{sec:data-aggregation},
one naturally arrives at a partition of one microblock per work group.
But even this can be further generalized, as one may work on more than
one microblock in each work group. We assume here that the work on
each work item is independent, but may depend on some preparation,
such as fetching matrix data into on-chip memory. This opens up a
number of possible avenues: One may\dots
\begin{itemize}
  \item process microblocks in parallel, adding more work items to 
    each work group, to achieve better usage of individual compute
    units.
  \item process microblocks in sequence, leaving the number of work
    items unchanged, but doing more work in each work item, thus
    amortizing preparation work.
  \item process multiple work items along with each other, reusing
    auxiliary data (such as matrices) that is already present in
    machine registers. We term this usage ``\emph{in-line parallel}''.
  \item use any combination of the above.
\end{itemize}
Figure \ref{fig:work-partition} illustrates these possibilities.

If a strategy is chosen that exploits the parallel processing of
multiple microblocks in one work group (the first option) above,
subtle questions of thread ordering arise that may influence the
number of local memory bank conflicts. In
\citep{kloeckner_nodal_2009}, we discuss one such question in more
detail.

Note that all combinations of parallel, sequential, and in-line
parallel do the same amount of work, and should, in theory, require
similar time to complete. In practice, this is not the case.  This
begs the question of how to decide between the numerous different
possibilities. It is of course possible to explore manually which
combination yields the best performance, but this is tedious,
error-prone, and needs to be repeated for nearly every change to the
hardware on which the code is run. This clearly undesirable, but a
potential solution is described in the next section.
\subsubsection{Using Run-Time Code Generation}
In \citep{kloeckner_pycuda_2009}, we discuss the numerous benefits of
being able to generate computational code immediately before it is
used, i.e. being able to perform C-level \emph{run-time code
generation}. We are delighted that OpenCL has this capability built
into its specification. We do note however that the feature can be
retrofitted onto CUDA through the use of the \emph{PyCUDA} Python
package. In our DG demonstration code, we already make simple use of
this facility, by using string substitution on the source code of our
kernels to make certain problem size parameters known to the compiler
at compilation time. This helps decrease register pressure and allows
the compiler to use a number of optimizations such as static loop
unrolling for loops whose trip counts are now known.

But this is far from the only benefit. Another immediate advantage is
the ability to perform automated tuning to answer questions such as
the one raised at the end of the last section, where individual
kernels can be generated to cover any number of code variants to be
tried. Once this has been accomplished, implementing automated tuning
can be as simple as looping over all variants and comparing timing
data for each. For larger search spaces, a more sophisticated strategy
might be desirable. This entire topic is discussed in much greater
detail in \citep{kloeckner_pycuda_2009}.
\subsubsection{Further Tuning Opportunities}
In the demonstration code, some inefficiency lies buried in the way
the surface fluxes are evaluated. Because the data required by the
surface evaluation grows as $O(N^{d-1})$, whereas the volume data's
size grows as $O(N^d)$, this is mainly felt at low-to-moderate $N$,
which are particularly relevant for practical purposes. 

First, the index data loaded into \textsf{idM} and \textsf{idP} has
significant redundancy and can easily be compressed by breaking it
down into element offsets added to one particular entry from a list of
subindex lists. This list of subindex lists is comparatively small and
has better odds of being able to reside in on-chip memory.

Second, data for faces lying opposite to each other is fetched
twice--once for each side--in our current implementation. Through a
blocking strategy (which, unfortunately, introduces significant
complexity) these redundant fetches can be avoided. The strategy is
discussed in detail in \citep{kloeckner_nodal_2009}.

The last opportunity for tuning we will discuss in this setting is the
use of multiple GPUs through MPI or Pthreads. Since only facial data
for flux computation needs to be exchanged between GPUs, such a code
is cheap in communications bandwidth and relatively easy to implement.
We will now turn our discussion here from opportunities for speed
increase to ways of making the technology more useful and more broadly
applicable.

\subsection{Adding Generality}

GPU-DG as demonstrated in the demonstration code in this chapter can
be extended in a number of ways to address a larger number of
application problems.

\begin{description}
\item[Three Dimensions] Perhaps the most gentle, but also the most
immediately necessary generalization is the use of three-dimensional
discretizations. The main complexity here lies in adapting the 
set-up code that generates matrices and computes mesh connectivity.
The GPU kernels only require mild modification, although a number
of complexity trade-offs change, requiring different tuning decisions.
It is further helpful to generate general, $n$-dimensional code from
a single source through run-time code generation, to reduce
the amount of code that needs to be maintained and debugged.

\item[Double Precision] We have found that for most engineering
problems, single precision calculations are more than sufficient.  We
do however acknowledge that some problems \emph{do} require double
precision, and our methods can be easily adapted to accommodate it.

\item[General Boundary Conditions] Our demonstration code only
provided facilities for a single (Dirichlet) boundary condition. In
nearly all practical problems, multiple boundary conditions (BCs) are
needed,
ranging from Neumann to absorbing BCs to even more complicated
conditions arising in fluid dynamics.

\item[General Linear Systems of Conservation Laws] In addition to more
general BCs, one obviously often wants to solve more general PDEs than
the 2D wave/Maxwell equation discussed here--this might include
Maxwell's equations in 3D or the equations of aeroacoustics. Again,
making these adaptations to the demonstration code is relatively
straightforward and mainly entails implementing a different local
differentiation operator along with a new flux expression.

\item[Nonlinear Systems of Conservation Laws] Once general
\emph{linear} systems are treated by GPU-DG, it is, conceptually, not a
very big step to also treat \emph{non-linear} systems of conservation
laws, such as Euler's equations of gas dynamics or the even more
complicated Navier-Stokes equations, potentially along with various
turbulence models. The good news is that the methods presented so far
again generalize seamlessly and work well for simple problems.

However, due to the subtle subject matter, a number of refinements of
the method may be required to successfully treat real application
problems.

The first issue revolves around the evaluation of non-linear terms on
a nodal grid and the aliasing error thus introduced into the method.
One possibility of addressing this is filtering, which is easily
implemented, but impacts accuracy. Another is over-integration using
quadrature and cubature rules to more accurately approximate the
integrals involved in the method. A more detailed discussion of these
subjects is beyond the scope of this chapter and can be found in
\citep{hesthaven_nodal_2007}.

Shocks, i.e. the spontaneous emergence of very steep gradients in the
solution, are another complication that only arises in nonlinear
problems. Some initial ideas on using GPU-DG in conjunction with
shock-laden flow computations are available in
\citep{kloeckner_viscous_2010}.

\item[Curved Geometries] While finite element methods already offer
much greater flexibility in the approximation of geometry than solvers
on structured grids, the demonstration solver shown here is restricted
to geometries that consist of straight surfaces. A cost-efficient way
to extend this solver to curved geometries is shown in
\citep{warburton_accelerating_2008}.

\item[Local Time Integration] Lastly, as the solvers described in
this chapter all employ explicit marching in time, time step
restrictions may become an issue if the mesh involves very small
elements. \citep[Chapter 8]{kloeckner_phd_2010} describes a number
of time stepping schemes that can help overcome this problem.
\end{description}

As we have seen, a simple solver employing discontinuous Galerkin
methods on the GPU can be written without much effort. On the other
hand, far more effort can be spent on performance tuning and
adaptation to more general problems. A free solver that implements
nearly all of the improvements described here is available at
\url{http://mathema.tician.de/software/hedge} under the GNU Public
License.

\section{Final Evaluation}
\ednote{This section can also be a multi-section section, presenting 
final results, evaluation, validation of domain relevant quality, 
benefits and conclusion.  
Please do mention any known limitations in addition to the benefits.}

\begin{figure}
  \centering
  \includegraphics[width=0.7\textwidth]{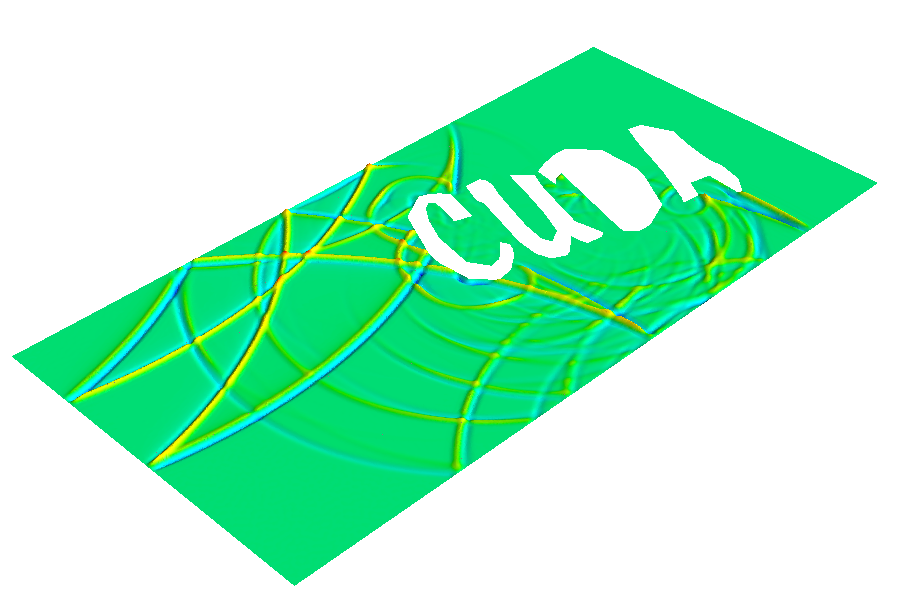}
  \caption{
    Screen shot of the demonstration solver showing a live snapshot of
    the simulation of a wave propagation problem on a moderately
    complex domain.
  }
  \label{fig:pydgeon-screenshot}
\end{figure}

In the present chapter, we have shown that, even with limited effort,
large performance gains are realizable for discontinuous Galerkin
methods using explicit time integration. Figure
\ref{fig:pydgeon-screenshot} shows a  live snapshot of the simulation
of a wave propagation problem on a moderately complex domain as shown
during run time by the solver if the
\texttt{mayavi2}\footnote{\url{http://code.enthought.com/projects/mayavi/}}
visualization package is installed.

\begin{figure}
  \subfigure[GFlops/s plotted vs. polynomial degree for the
  2D demonstration solver described in this chapter.]{
    \label{fig:order-gflops-pydgeon}
    \includegraphics[width=0.5\textwidth]{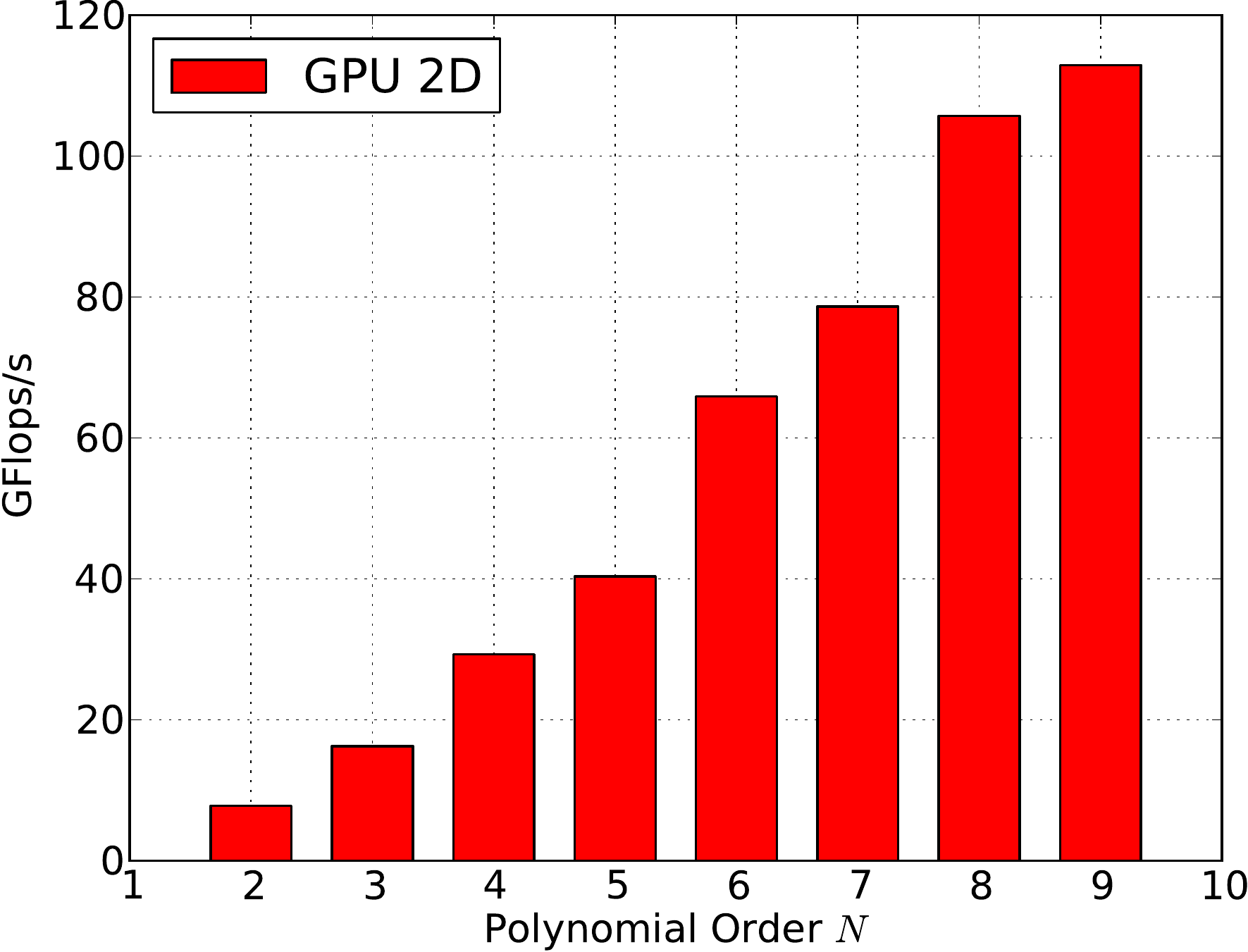}
  }
  \hspace{.4cm}
  \subfigure[GFlops/s plotted vs. polynomial degree for the
  3D DG solver hedge.]{
    \label{fig:order-gflops-hedge}
    \includegraphics[width=0.5\textwidth]{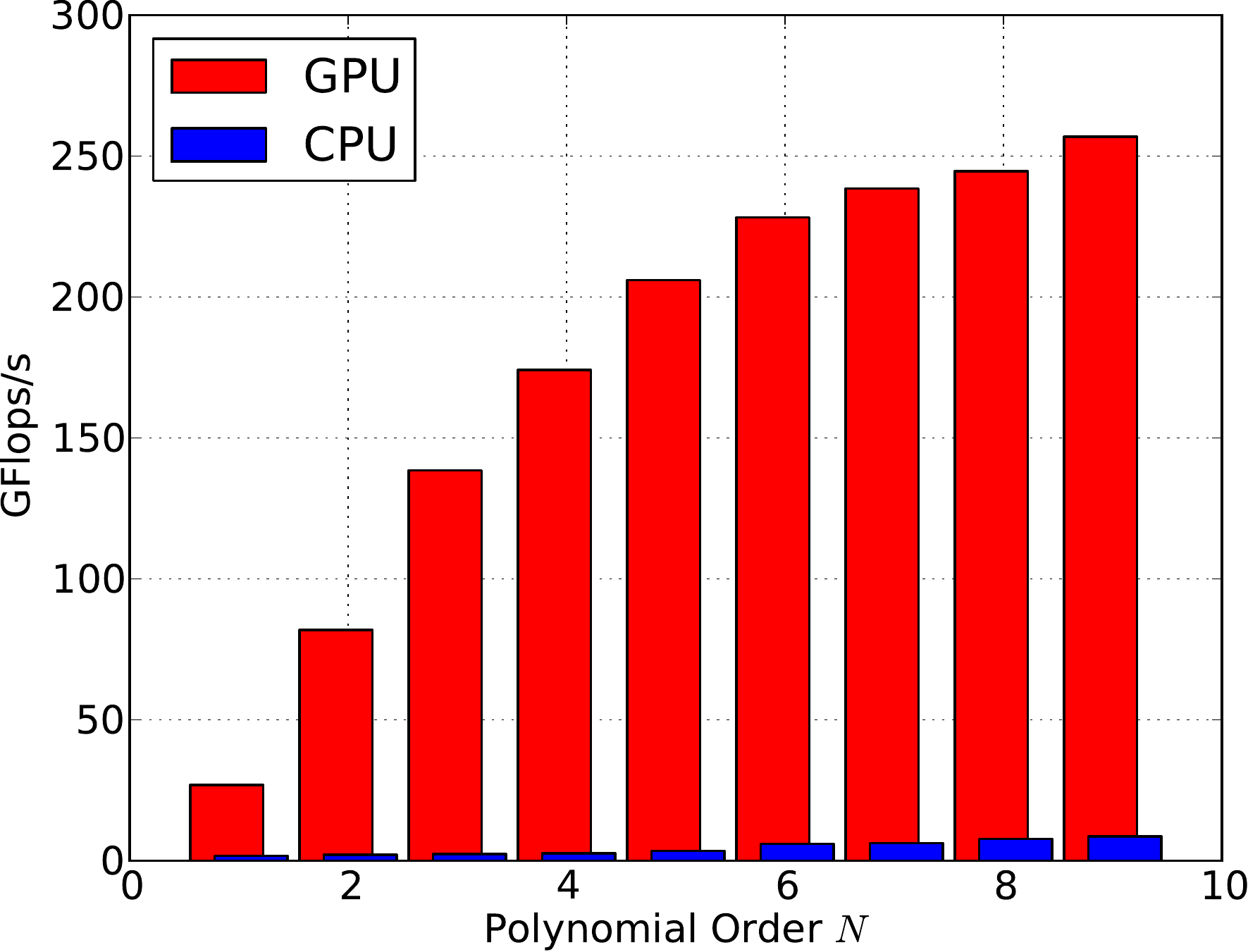}
  }
  \caption{
    Performance figures for the demonstration solver (2D) and the
    full-featured (3D) solver `hedge', both executing solvers
    for the Maxwell problem on an Nvidia GTX 295 in single precision.
  }
\end{figure}

Figure \ref{fig:order-gflops-pydgeon}
shows the performance of the demonstration solver developed here and
compares it to the performance of hedge, our (freely available) 
production solver. Note that these graphs should not be compared
directly, as one shows the result of a 2D simulation, while the other
portrays the performance of a three-dimensional computation. With some
care, a few observations can be made however.

It is possible to observe that, as the element-local matrices grow
with the third power of the degree in three dimensions (as opposed to
the second power in 2D), they constitute a larger part of the
computation and contribute many more floating point operations,
leading to higher performance. For the same reason, performance
increases as the polynomial degree $N$ increases.

Many of the tuning ideas described in Section
\ref{sec:improving-performance} (such as microblocking and
matrix-in-local) are designed to help performance in the case of lower
$N$, and hence, smaller matrices. Without comparing absolute numbers,
we observe that the initial performance increase at low $N$ is much
faster in the production solver than in the demonstration solver. We
attribute this to the implementation of these extra strategies, that
lead to markedly better performance at moderate $N$. In \citep{kloeckner_nodal_2009},
we also briefly study the individual and combined effects of a few of
these performance optimizations.

As a final observation, we would like to remark that the performance
obtained here is very close to to the performance obtained in a
C-based OpenCL solver written for the same problem--the use of Python
as an implementation language does not hamper the speed of the solver
at all. In particular, this facilitates a very logical splitting of
computational software into performance-critical, low-level parts
written in OpenCL C, and performance-uncritical set-up and
administrative parts written in Python.

\section{Future Directions}
\ednote{This is perhaps the smallest section with some suggestions on future directions.}


We have shown that, using our strategies, high-order DG methods can
reach double-digit percentages of published theoretical peak
performance values for the hardware under consideration. This
computational speed translates directly into an increase of the size
of the problem that can be reasonably treated using these methods.  A
single compute device can now do work that previously required a
roomful of computing hardware--even using the simplistic
implementation demonstrated here.

It is our stated goal to further broaden the usefulness of the method
through continued investigation of the treatment of nonlinear
problems, improved time integration characteristics, and coupling to
other discretizations to optimally exploit the characteristics of
each. GPUs present a rare opportunity, and it is fortuitous that a
method like DG, which is known for highly accurate solutions, can
benefit so tremendously from this computational advance.

\subsection*{Acknowledgments}

We would like to thank Xueyu Zhu, who performed the initial Python
port of the Matlab codes from which the demonstration solver is
derived. 

TW acknowledges the support of AFOSR under grant number
FA9550-05-1-0473 and of the National Science Foundation under grant
number DMS 0810187. JSH was partially supported by AFOSR, NSF, and
DOE.  AK's research was partially funded by AFOSR under contract
number FA9550-07-1-0422, through the AFOSR/NSSEFF Program Award
FA9550-10-1-0180 and also under contract DEFG0288ER25053 by the
Department of Energy.  The opinions expressed are the views of the
authors.  They do not necessarily reflect the official position of the
funding agencies.

\bibliographystyle{plainnat}
\bibliography{main}

\end{document}